\UseRawInputEncoding
\documentclass[12pt,preprint]{aastex}
\usepackage{natbib}
\usepackage{longtable}
\usepackage{rotating}
\usepackage[usenames,dvipsnames]{color}
\usepackage{ulem}

\def\cm2{cm$^{-2}$}

\def\nh3{NH$_3$}
\def\n2h{N$_2$H$^+$}

\def\13co{$^{13}$CO}
\def\c18o{C$^{18}$O}
\def\hc3n{HC$_3$N}
\def\h2{H$_2$}
\def\nh{n(H$_2$)}

\begin{document}

\slugcomment{}

\shorttitle{The ITRF coordinates of the spherical center of FAST} \shortauthors{Qian, Yue}

%%-3Sep2015
\title{The ITRF coordinates of the spherical center of FAST}
%\title{HI Properties of four IMBH Hosting Galaxies (Report of A2935 run)}
\author{ Lei Qian\altaffilmark{1}, Youling Yue\altaffilmark{1}}
\altaffiltext{1} {National Astronomical Observatories, Chinese Academy of
Sciences, Beijing 100101, China}

\begin{abstract}
The ITRF coordinates of the spherical center of the Five-hundred-meter Aperture Spherical radio Telescope (FAST) are $(X,Y,Z)=(-1668557.2070983793,$ $5506838.5266271923, 2744934.9655897617)$.
\end{abstract}

%%\keywords{ISM: clouds --- ISM: molecules --- Galaxis: active} %- 3Sep2015
%% -I"m not sure if \hi\ is a keyword. --

\section*{}

Five-hundred-meter Aperture Spherical radio Telescope \citep[FAST,][]{2011IJMPD..20..989N} is the largest single-dish radio telescope in the world, starting to achieve in the studies of pulsars and fast radio bursts \citep{QIAN2020100053}.

The ITRF coordinates of a fixed point of an observatory is required for pulsar search and pulsar timing software. For the Five-hundred-meter Aperture Spherical radio Telescope (FAST), the fixed point can be taken as the spherical center of the the reflector. In using various software, we find all kinds of inconsistent coordinates of FAST. Here we try to nail it down.

In the WGS-84 coordinate, the longitude, latitude, and altitude of the spherical center of FAST are $\lambda=106^\circ 51' 24.000740"$, $\phi=25^\circ 39'10.626537"$, and $h=1110.028801 {\ \rm m}$, respectively (from FAST technical report). The ellipsoids of ITRF and WGS-84 are almost the same, with a difference less than 1 meter. Here we use the WGS-84 ellipsoid. The transform from $(\lambda, \phi, h)$ to $(X, Y, Z)$is
\begin{eqnarray*}
X&=&(N+h)\cos\phi\cos\lambda\\
Y&=&(N+h)\cos\phi\sin\lambda\\
Z&=&[N(1-e^2)+h]\sin\phi\\
\end{eqnarray*}
where $N=a/\sqrt{1-e^2\sin^2\phi}$, the semi-major $a=6378137 {\ \rm m}$, semi-minor

\noindent $b=6356752.3142451795 {\ \rm m}$, $e^2=\frac{a^2-b^2}{a^2}$¡£

It is straight forward to get $(X,Y,Z)=(-1668557.2070983793,$

\noindent $5506838.5266271923, 2744934.9655897617)$. This value is consistent with that used by PINT, $(X,Y,Z)=(-1668557.0, 5506838.0, 2744934.0)$, which is provided by Youling Yue, one of the authors. We think the FAST coordinate used by PINT is OK. More accurate coordinates will be obtained with pulsar timing and VLBI observations.

%\bibliography{msbib}{}
%\bibliographystyle{apj}

\end{document}